\documentclass{article}
\date{}
\usepackage[english]{babel}

\usepackage[letterpaper,top=2cm,bottom=2.5cm,left=3cm,right=3cm,marginparwidth=1.75cm]{geometry}

\usepackage{amsmath}
\usepackage{graphicx}
\usepackage[table,xcdraw]{xcolor}
\usepackage{caption}
\usepackage{subcaption}
\usepackage[colorlinks=true, allcolors=blue]{hyperref}

\title{Ethical Framework for Harnessing the Power of AI in Healthcare and Beyond}

\usepackage{authblk}

\author{Sidra Nasir}
\author{Rizwan Ahmed Khan\thanks{corresponding author: Rizwan.khan@shu.edu.pk}}
\author{Samita Bai}
\affil{Department of Computer Science, Faculty of Information Technology, Salim Habib University, Karachi, Pakistan}

\begin{document}
\maketitle

\begin{abstract}
In the past decade, the deployment of deep learning (Artificial Intelligence (AI)) methods has become pervasive across a spectrum of real-world applications, often in safety-critical contexts. This comprehensive research article rigorously investigates the ethical dimensions intricately linked to the rapid evolution of AI technologies, with a particular focus on the healthcare domain. Delving deeply, it explores a multitude of facets including transparency, adept data management, human oversight, educational imperatives, and international collaboration within the realm of AI advancement. Central to this article is the proposition of a conscientious AI framework, meticulously crafted to accentuate values of transparency, equity, answerability, and a human-centric orientation. The second contribution of the article is the in-depth and thorough discussion of the limitations inherent to AI systems. It astutely identifies potential biases and the intricate challenges of navigating multifaceted contexts. Lastly, the article unequivocally accentuates the pressing need for globally standardized AI ethics principles and frameworks. Simultaneously, it aptly illustrates the adaptability of the ethical framework proposed herein, positioned skillfully to surmount emergent challenges.

\end{abstract}

\section{Introduction}
Artificial intelligence (AI) has shown immense promise in transforming healthcare through the application of advanced technologies like machine learning (ML) and deep learning (DL) \cite{bini2018artificial, dimiduk2018perspectives}. These techniques enable the processing and analysis of vast amounts of medical data, leading to improved patient care through pattern identification and valuable insights extraction. However, despite these advancements, certain limitations hinder the widespread adoption and integration of AI into clinical practice, particularly in terms of explainability and interpretability \cite{belle2021principles, DBLP:journals/corr/abs-1807-06722}.

Unlike traditional computer programs that follow a predefined set of instructions, ML and DL models learn from data to identify patterns and make predictions \cite{kocher2021machine}. This self-programming ability, coupled with the immense computing power of modern computers, allows models to process vast amounts of healthcare data quickly. While simple tasks can be automated with conventional programs, it is particularly suited for complex cognitive tasks in healthcare, such as natural language translation, predictive maintenance of medical equipment, and large-scale image analysis for object recognition \cite{collins2019reporting}. In low-stakes applications, these models can autonomously make decisions. For example, they can assist healthcare professionals in diagnosing diseases or recommend personalized treatment plans. In high-stakes situations, these models can significantly enhance the efficiency and accuracy of human decision-making \cite{lee2021application}. However, it is crucial to ensure that these systems support human users and operators who bear the final responsibility for decision-making. The models should also act as "super-assistants," providing an additional layer of insight and analysis to aid healthcare professionals in their decision-making process \cite{nauta2023explainable}. Yet with benefits, limitations of AI can not be ignored \cite{burger2023hybrid}.

The size and complexity of models, particularly deep neural networks (DNNs), have increased in pursuit of better predictive performance. However, there has been criticism of solely focusing on predictive accuracy. Relying on large, opaque models raises concerns about the lack of transparency in decision-making processes. In the healthcare context, the lack of interpretability \cite{zhang2021survey} and explainability \cite{rosenfeld2019explainability} in ML and DL models can lead to ethical issues and a loss of trust \cite{ferrara2023fairness, holzinger2019causability}. Understanding why a particular decision is made is just as crucial as knowing what the decision is. Insufficient interpretability hinders the widespread and responsible adoption of ML, especially in high-stakes domains where transparency and accountability are paramount. However, addressing the issue of interpretability can open doors to valuable future applications of ML in healthcare \cite{ghassemi2021false}.

Therefore, there is a growing need for explanations that help healthcare professionals understand and trust AI systems. Explanations can provide insights into the reasoning behind ML predictions, ensuring that the learned patterns align with expectations and intentions. Transparent and interpretable ML models are essential for fostering trust, ethical decision-making, and responsible adoption of AI in healthcare. Additionally, the lack of transparency in traditional AI algorithms, particularly DL models, leads to them being seen as enigmatic systems with concealed internal operations. This lack of clarity presents notable obstacles when it comes to establishing trust and gaining approval from healthcare professionals, patients, and regulatory bodies. However, these challenges can be overcome by providing explanations and interpretations that shed light on the inner workings and decision-making processes employed by the model to achieve its results \cite{DBLP:journals/corr/abs-1807-06722}.

\begin{figure}
\centering
\includegraphics[width=1\linewidth]{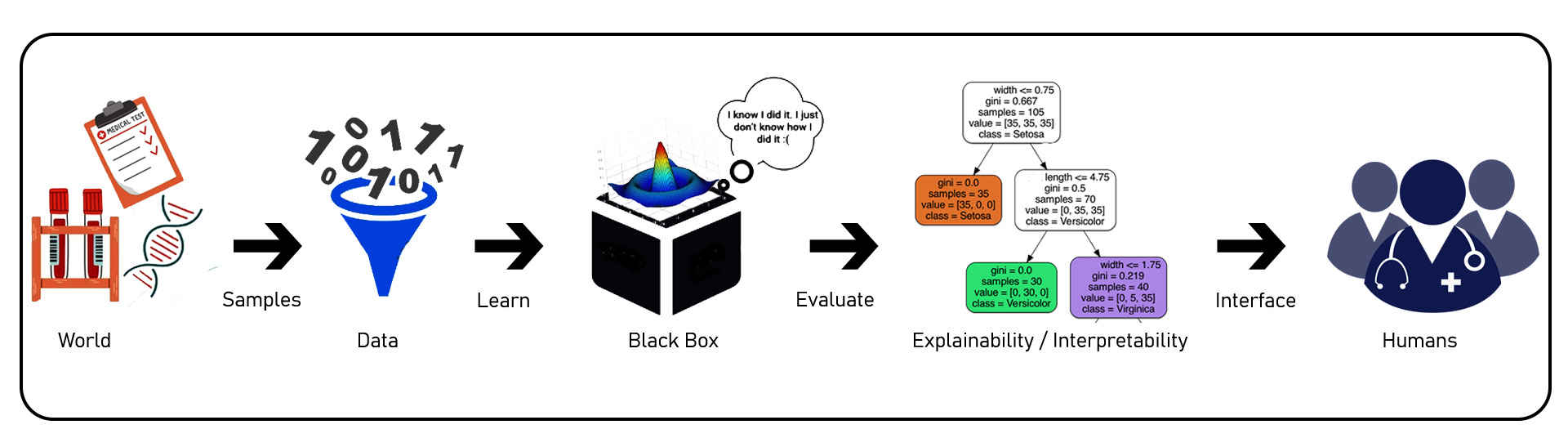}
\caption{\label{fig:1.1}Machine Learning from data collection to its interpretability and explainability to humans}
\end{figure}

\begin{figure}
\centering
\includegraphics[width=1\linewidth]{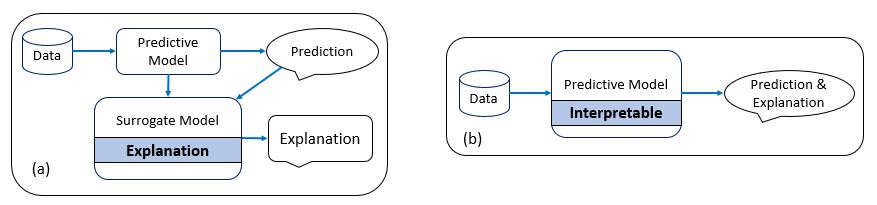}
\caption{\label{fig:1.2} (a)Explainabilty vs (b) Interpretability.}
\end{figure}

Interpretability refers to the ability to understand the inner workings or mechanics of a ML model, particularly in the context of predicting temperature over time in a normal regime. It allows us to gain insights into how the model generates its predictions without necessarily understanding the underlying reasons or causality behind those predictions \cite{doshi2017towards, lipton2018mythos}. In other words, interpretability is a desirable component for explainability, but not all interpretable models are explainable \cite{guidotti2018survey}. Additionally, it deals with counterfactual cases, enabling us to examine what would have changed if certain features or values had been different. It aims to develop a comprehensive understanding of the ML system, including both observed and unobserved factors, towards creating a global theory. The black-box nature of AI models restricts the transparency of their outputs, making it difficult to comprehend the reasoning behind their decisions \cite{lakkaraju2017interpretable} as illustrated in Figure \ref{fig:1.1}. Nevertheless, it focuses on comprehending the operational aspects of the model while explainability goes beyond interpretability by encompassing the ability to address questions related to the model's behavior in scenarios involving new data. It involves understanding the consequences of specific actions or changes in the model's predictions as shown in Figure \ref{fig:1.2}. In healthcare, where decisions can have life-and-death consequences, it is crucial to have a clear understanding of how AI algorithms arrive at their recommendations. Without explainability, healthcare professionals may find it challenging to trust and validate the outputs of AI systems, leading to resistance to adopting these technologies \cite{gilpin2018explaining}.

Moreover, the lack of interpretability hampers the ability to identify and mitigate biases in the data or algorithms used by AI models. Biased data or biased model predictions can have serious consequences in healthcare, resulting in inaccurate diagnoses, suboptimal treatments, and potential harm to the patients. Without the ability to understand and explain the reasoning behind AI-generated recommendations, it becomes difficult to identify and address these biases effectively. The absence of explainability in AI systems also raises ethical concerns. In healthcare, it is essential to provide justifications and explanations for decisions that impact patients' well-being. When AI systems cannot provide transparent and interpretable explanations for their recommendations, it becomes challenging to understand and justify the basis of these decisions. This lack of transparency may result in ethical dilemmas and undermine the accountability of AI systems in healthcare \cite{vandewiele2016genesim}. For example, in \cite{lee2023benefits} the authors discussed that, 
GPT-4 as an AI chatbot for medicine has benefits such as providing accessible medical information, availability round-the-clock, reducing healthcare professionals' workload, and aiding patient education and language translation. However, it has certain limitations including the inability to understand the full context, lack of clinical judgment, and reliance on training data. Risks involve potential misinformation, over-reliance on AI, ethical and legal concerns, and the absence of emotional support. It should be used cautiously to complement human expertise rather than replace it entirely. Hence, the question arises; What is an intelligent system, and what does it entails? AI involves the pursuit of imbuing machines with intelligence, wherein intelligence refers to the capability of an entity to operate effectively and with foresight within its surroundings \cite{nilsson2009quest}.

To overcome these limitations, the development and adoption of Explainable Artificial Intelligence (XAI)  techniques in healthcare are imperative. XAI aims to enhance the transparency and interpretability of AI models, providing insights into the decision-making process and enabling clinicians to understand the reasoning behind AI-generated recommendations \cite{adadi2020explainable}.

Various approaches and techniques have been proposed to achieve explainability in AI models. One approach is to use interpretable algorithms that provide explicit rules or representations of the decision-making process. These models offer transparency by explicitly showing how specific inputs lead to certain outputs \cite{lin2021you}. Another approach is to develop post-hoc explainability methods that provide explanations after the AI model has made its predictions. These methods shed light on the internal workings of black-box models by generating explanations as feature importance, saliency maps, or textual descriptions. Techniques such as LIME \cite{ribeiro2016should} and SHAP \cite{lundberg2017unified} are examples of post-hoc explainability methods commonly used in healthcare.

Furthermore, advancements in DL have led to the emergence of techniques specifically designed to enhance explainability in DNNs. Methods like attention mechanisms and gradient-based techniques such as GradCAM \cite{selvaraju2017grad} provide insights into the areas of focus and decision-making process of the models \cite{arrieta2020explainable}. The benefits of XAI in healthcare are numerous. By integrating transparency and interpretability into AI models, these systems can instill trust and confidence among healthcare professionals and patients. Clinicians can better understand and validate the outputs of AI algorithms, leading to increased adoption and collaboration between human experts and AI systems. Moreover, it can also facilitate regulatory compliance by enabling audits and assessments of AI models to ensure fairness, accountability, and compliance with ethical standards \cite{amugongo2023operationalising, markus2021role}. In medical imaging, for example, radiologists can benefit from such systems that provide explanations for their findings, aiding in the detection and diagnosis of diseases. Clinical decision support systems (CDSS) can also provide justifications for treatment recommendations, empowering healthcare professionals to make informed decisions based on a better understanding of the underlying reasoning. To address these limitations, there is a growing need for the development and adoption of XAI techniques in healthcare \cite{loh2022application}. By fostering explainability, AI systems can instill trust, improve regulatory compliance, and enhance collaboration between human experts and AI algorithms as they provide insights into the decision-making process and enable clinicians to understand the reasoning behind AI-generated recommendations.

The rapid evolution of AI technologies has outpaced the formulation of comprehensive ethical guidelines, raising concerns about the potential misuse and unintended consequences of these powerful tools \cite{o2019legal, eitel2021beyond}. In the realm of healthcare, these concerns encompass a wide spectrum of issues, including data privacy and security, algorithmic bias, accountability, and the potential dehumanization of patient care. The design and implementation of AI systems should be underpinned by ethical considerations that prioritize transparency, fairness, and the responsible handling of patient data. The need for an ethical framework becomes even more pronounced as AI is poised to extend its influence beyond healthcare into various domains, such as education, transportation, and governance \cite{peters2020responsible}.

While the benefits of AI in healthcare are evident, a critical examination of its ethical implications is crucial to avoid unintended negative consequences. The potential for biases embedded within algorithms to exacerbate health disparities and perpetuate social injustices underscores the necessity of vigilance in algorithm development. Striking a balance between innovation and ethical reflection necessitates interdisciplinary collaboration involving AI developers, healthcare professionals, ethicists, policymakers, and patients \cite{morley2020ethically}. By fostering open dialogues among these stakeholders, it becomes possible to create a dynamic ethical framework that adapts to the evolving landscape of AI applications and promotes responsible AI development \cite{naik2022legal}. The application of AI in healthcare brings forth several limitations that need to be addressed for its successful integration into clinical practice \cite{rasheed2022explainable}. 

This paper comprehensively examines the ethical dimensions of AI development, with a focus on healthcare. It outlines a framework emphasizing transparency, fairness, and human-centricity while delving into the challenges of AI integration in healthcare. The paper acknowledges AI's limitations, highlighting the need for ongoing research, preserving human empathy, and fostering responsible AI practices. It underscores the significance of education, advocates for global ethical standards, and emphasizes the adaptive nature of the framework. Ultimately, this paper offers a profound exploration of AI ethics, guiding its responsible advancement amidst complex societal considerations.

\section{Black Box and Lack of Transparency}
In the field of AI, black box models refer to ML algorithms or DNNs that produce accurate predictions or decisions but lack explainability. These models can provide remarkable results in tasks such as image recognition, natural language processing, or autonomous driving. However, the inner workings of these models are often complex and challenging to explain in human terms \cite{bathaee2017artificial, makhni2021artificial}. This lack of interpretability can limit our ability to understand why certain decisions are made, leading to potential biases, discrimination, or errors that go unnoticed \cite{ribeiro2016should}.

ML has the ability to develop algorithms for diagnosis, prognosis, and outcome prediction by utilizing various features. However, the lack of transparency regarding the reasoning/understanding behind these algorithms creates a "black box" situation, where the inputs and outputs are not easily explainable. While this may be acceptable in certain fields such as business decisions or human behavioral studies, it becomes problematic in clinical management, where decisions often involve critical, life-or-death situations \cite{ghassemi2020review}.

Clinicians facing such scenarios are understandably concerned about trusting the model's appropriateness for their patients, its accuracy in guiding clinical judgment, and its alignment with existing knowledge of human disease. There are three key aspects to address: trust, consistency, and explanation \cite{zihni2020opening}.

Conventional medical decisions are based on a thorough understanding of pathophysiological mechanisms, supported by cell-based experiments, genomic and metagenomic analysis, animal studies, histopathological observations, clinical trials, and cohort observations. Evidence-based medicine has long been the gold standard for treatment strategies. The questions which arise in \cite{london2019artificial, poon2021opening} are as follows:
\begin{enumerate}
  \item Can clinicians make vital decisions without grasping the fundamental reasoning behind them?
  \item Can patients undergo treatment, surgery, or cancer therapy without fully understanding the rationale behind the chosen course of action?
  \item Do doctors have the capability to adequately elucidate to patients why they prefer a specific treatment over other alternatives?
  \item How do we convince clinicians and scientists that ML models can excel in clinical decision-making beyond evidence-based diagnosis and treatment? 
  \item The lack of trust in ML algorithms poses a considerable challenge in implementing AI medicine. In case of adverse outcomes, who bears responsibility for the resulting consequences?
\end{enumerate}

Many AI algorithms, such as DNNs, operate as black-box models, making it challenging to understand the reasoning behind their decisions. While these models may produce accurate results, their lack of transparency raises concerns about trust, interpretability, and accountability. The inability to explain how decisions are made hinders the acceptance and adoption of AI in critical healthcare scenarios \cite{DBLP:journals/corr/abs-1807-06722}.

\begin{figure}
\centering
\includegraphics[width=1\linewidth]{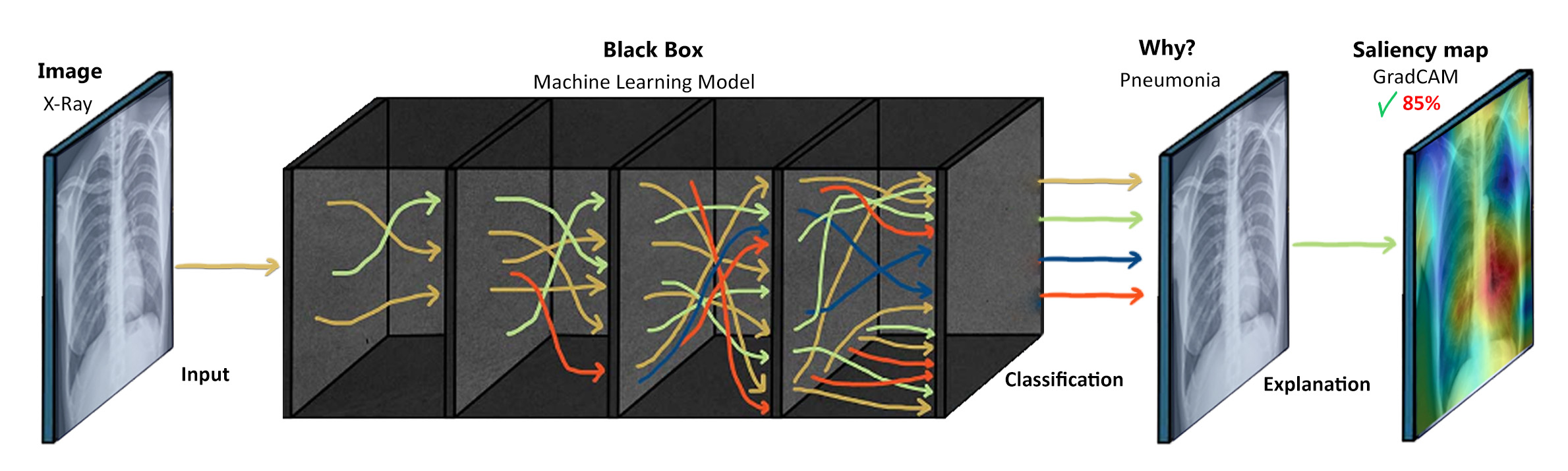}
\caption{\label{fig:1.10}Example of explainability of deep learning model: Pneumonia detection using GradCAM}
\end{figure}

To illustrate the challenges posed by black-box models in healthcare, let's consider an example of using DNNs for diagnosing diseases from medical images, such as chest x-rays \cite{ren2021interpretable}. DL models, particularly convolutional neural networks (CNNs), have shown remarkable performance in detecting various diseases from images accurately. However, understanding how these models arrive at their predictions is often elusive due to their black-box nature. In the case of chest x-ray diagnosis, a DL model is trained on a large dataset of labeled images to learn patterns and features indicative of different diseases, such as pneumonia \cite{ali2023pneumonia} or lung cancer. The model goes through numerous iterations, adjusting its internal parameters until it can accurately classify the images based on the training data. Once trained, the black-box nature of the model becomes apparent. When a new chest x-ray image is presented to the model for diagnosis, it produces a prediction, such as "pneumonia present" or "no pneumonia." However, the model does not provide explicit information about which regions or features in the image led to that specific prediction. This nature of results is not acceptable by clinicians, hence a surrogate model is applied to explain and verify the validity of the results obtained as exhibited in Figure \ref{fig:1.10}.

Similarly, breast cancer is a prevalent disease, and early detection is crucial for successful treatment \cite{shah2022artificial}, and CNNs have also shown promising results in accurately identifying malignant tumors in mammogram images \cite{sajid2023breast}.

However, their black-box nature presents challenges in understanding their decision-making process. When training a DL model for breast cancer detection, a large dataset of labeled mammogram images is used to train the model to recognize patterns indicative of cancerous lesions. The model learns to analyze various features, such as the shape, texture, and density of potential tumors, in order to make predictions. Once trained, the black-box nature of the model becomes apparent during the inference phase. When a new mammogram image is provided to the model for evaluation, it produces a prediction indicating whether a malignant tumor is present or not. However, it does not explicitly provide information on which regions or characteristics of the mammogram contributed to that particular prediction \cite{karatza2021interpretability}. 

Moreover, the same can be observed in AI technologies that have demonstrated their potential in addressing various aspects of autism. ML algorithms can analyze large datasets to identify patterns and markers associated with Autism spectrum disorder (ASD), aiding in early diagnosis and personalized treatment plans. AI-driven virtual reality environments and social communication tools offer innovative platforms to develop and practice social skills in a controlled and supportive setting \cite{jaliaawala2020can}. ASD is a highly heterogeneous condition, and interventions need to be tailored to an individual's unique needs. Black box models might struggle to explain why a particular intervention is recommended for a specific individual, making it challenging to adapt and personalize interventions effectively \cite{sharif2022novel}.

 This lack of interpretability can be a significant barrier to the adoption of AI in healthcare. Clinicians need to understand the rationale behind the model's decision in order to validate its predictions and make informed treatment decisions. They may want to identify specific areas of concern within the images that influenced the model's classification. While patients may also desire transparency and explanations for the model's predictions to gain confidence in the diagnosis and treatment recommendations \cite{srinivasu2022blackbox}. Additionally, patients may feel uneasy about relying on AI systems if they cannot comprehend how the diagnosis was made. They may have concerns about the transparency of the process, the reliability of the model, and the potential biases in the data used for training \cite{arcelus2007integration}. 
 
\section{Bias and Fairness}
\begin{figure}[ht]
\centering
\includegraphics[width=1\linewidth]{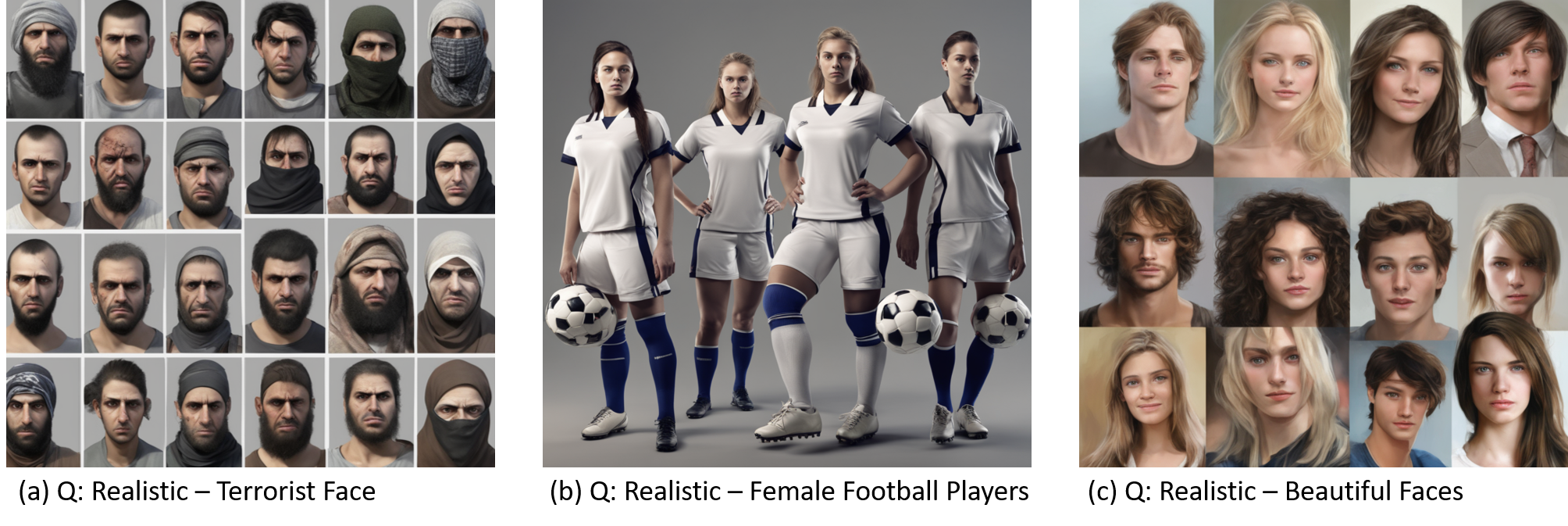}
\caption{\label{fig:1.3}Machine Learning systems (Playground AI) exhibiting gender and racial bias}
\end{figure}

The increasing adoption of DL and AI in healthcare holds significant promise for improving diagnostics, treatment, and healthcare delivery. However, the presence of bias regarding AI presents substantial challenges to achieving fairness in healthcare. Bias can be categorized as statistical or social, and it can lead to unequal outcomes and hinder equitable healthcare provision \cite{hendriksen2013diagnostic, nakayama2022global}. For instance, \cite{obermeyer2019dissecting} highlighted the significance of this issue through their study, which revealed racial bias in a commonly used algorithm in the U.S. healthcare system. The algorithm relied on healthcare spending to determine patients requiring extra medical attention. While spending might serve as a reasonable indicator of illness severity and the need for additional care, the authors showed that implementing the algorithm would result in a reduction of over 50\% in the number of Black patients identified as needing extra care. Similar racial and gender bias can be observed in an online AI image creator tool i.e. Playground AI \cite{mousavi2020ai} as exemplified in Figure \ref{fig:1.3}.

Data limitations pose a critical challenge in healthcare AI, as biased datasets can result in biased algorithmic outputs. Statistical bias occurs when a dataset does not accurately represent the true distribution of a population, potentially leading to algorithmic outputs that deviate from actual estimates. In the biomedical field, where large datasets are crucial, advanced information extraction methods are required to handle the complexity of the data. However, if the dataset used to train AI algorithms is not diverse and representative, it can introduce statistical bias and compromise the accuracy and generalizability of the algorithms. A notable example can be seen in cardiology \cite{maserejian2009disparities}, where heart attacks are frequently misdiagnosed in women. Prediction models for cardiovascular disease that claim to predict heart attacks several years in advance are often trained on datasets predominantly comprising male subjects \cite{ehz592}. However, cardiovascular disease manifests differently in men and women, and an algorithm primarily trained on male data may not accurately diagnose female patients \cite{jameson2018harrison}.

Social bias in healthcare AI refers to inequities resulting in sub-optimal outcomes for specific groups within the population. This type of bias is particularly concerning, as it unintentionally discriminates against vulnerable groups, perpetuating disparities \cite{davenport2019potential,celi2022sources}. For example, in dermatology, CNNs used to classify skin lesions are often trained on datasets predominantly composed of samples from patients with fair skin tones. This limited representation of patients with darker skin tones and its variations leads to lower diagnostic accuracy when these networks are tested with images of patients with darker skin tones. This disparity is especially problematic given the higher mortality rate for melanoma in patients with darker skin tones compared to Caucasian patients \cite{BRINKER201911}. Thus, leading to misdiagnosis and sub-optimal care.

The impact of data bias is not only limited to specific medical domains but extends across various areas of healthcare. For example, automated sleep scoring algorithms trained on data from young and healthy individuals struggle to accurately diagnose sleep disorders in older patients. Despite progress in developing these algorithms, achieving high performance in clinical routines remains challenging. Cognitive biases and the lack of diversity in the training data contribute to inter- and intra-scoring disagreement, posing limitations \cite{FIORILLO2019101204}.

Moreover in drug development and clinical trials, bias in AI algorithms can have significant consequences. Clinical trials often have predominantly male participants from limited age groups and similar ethnic backgrounds, leading to gender and ethnic biases. These biases during the pre-clinical stages can impact how women respond to newly developed drugs. This means that the results obtained from these early studies could influence the datasets used to train AI algorithms \cite{oh2015diversity, chandak2020using}.

While data bias is a critical concern, it is important to note that bias in clinical datasets is not the sole source of bias in healthcare AI. Researchers and clinicians can inadvertently introduce unconscious judgments and biases into their research, which may manifest in biased AI algorithms. These biases can further perpetuate health inequities and hinder the potential benefits of AI in healthcare. Fairness is fundamental, its principles include fair treatment for all individuals, group fairness (avoiding group-based disparities), statistical parity (equalizing outcomes across groups), equal opportunity (balanced true positive rates), and fairness through unawareness (ignoring certain attributes). Achieving fairness requires addressing biased training data, navigating complex interactions, and making trade-offs between fairness and accuracy \cite{ueda2023fairness}.

Therefore, data can be susceptible to various types of bias, which can significantly impact healthcare outcomes while fairness in AI focuses on ensuring that the outcomes and decisions produced by AI systems do not discriminate against specific groups or individuals based on their inherent characteristics, such as race, gender, age, or socioeconomic status. The goal is to mitigate bias and achieve equitable treatment across diverse demographics.

\begin{table}[ht]
\resizebox{\textwidth}{!}{%
\begin{tabular}{|l|ll|ll|ll|ll|ll|ll|
>{\columncolor[HTML]{D9D9D9}}l |}
\hline
DATASET & \multicolumn{2}{l|}{\cellcolor[HTML]{FFF2CC}I} & \multicolumn{2}{l|}{\cellcolor[HTML]{FCE5CD}II} & \multicolumn{2}{l|}{\cellcolor[HTML]{F9CB9C}III} & \multicolumn{2}{l|}{\cellcolor[HTML]{E69138}IV} & \multicolumn{2}{l|}{\cellcolor[HTML]{783F04}V} & \multicolumn{2}{l|}{\cellcolor[HTML]{5B0F00}VI} & TOTAL \\ \hline
PPB \cite{buolamwini2018gender}     & \multicolumn{1}{l|}{6.22\%}        & 79        & \multicolumn{1}{l|}{34.02\%}       & 432        & \multicolumn{1}{l|}{12.83\%}        & 163        & \multicolumn{1}{l|}{4.25\%}         & 54        & \multicolumn{1}{l|}{18.27\%}       & 232       & \multicolumn{1}{l|}{24.41\%}        & 310       & 1270  \\ \hline
IJB-A \cite{klare2015pushing}  & \multicolumn{1}{l|}{2.60\%}        & 13        & \multicolumn{1}{l|}{33.00\%}       & 165        & \multicolumn{1}{l|}{44.00\%}        & 220        & \multicolumn{1}{l|}{12.60\%}        & 63        & \multicolumn{1}{l|}{2.00\%}        & 10        & \multicolumn{1}{l|}{5.80\%}         & 29        & 500   \\ \hline
Adience \cite {levi2015age} & \multicolumn{1}{l|}{3.14\%}        & 69        & \multicolumn{1}{l|}{60.94\%}       & 1337       & \multicolumn{1}{l|}{22.15\%}        & 486        & \multicolumn{1}{l|}{8.07\%}         & 177       & \multicolumn{1}{l|}{4.38\%}        & 96        & \multicolumn{1}{l|}{1.32\%}         & 29        & 2194  \\ \hline
\end{tabular}%
}
\caption{Skin tone drive-bias exhibited in different datasets \cite{buolamwini2018gender}}
\label{tab:data-bias1}
\end{table}

\subsection{Data-Driven Bias}
Despite the independent process of analysis conducted by ML algorithms, the potential for bias remains present. The data bias could stem from either the dataset employed for data analysis or the data itself, such as datasets that lack diversity. For instance, in a study conducted by \cite{obermeyer2019dissecting} and colleagues, racial bias was identified in the initial data, leading to the erroneous conclusion that Black patients are more medically intricate and costly to the healthcare system compared to White patients. This oversight occurred because the model failed to consider issues of access. The subsequent problem has been glorified in \cite{buolamwini2018gender}, and statistics are exhibited in Table \ref{tab:data-bias1}. Similarly, Amazon's attempt to create an AI-driven tool for recruitment faced problems as the algorithm exhibited a negative bias against women due to the predominantly male-oriented data used for training purposes \cite{topol2019high}.

\emph{Measurement bias:} occurs when proxy variables are used to measure certain features, resulting in inaccurate assessments. For instance, in healthcare, relying solely on body mass index (BMI) as a measure of overall health can introduce measurement bias, as BMI may not accurately capture individual variations in body composition and health risks \cite{suresh2021framework}.

\emph{Omitted variable bias:} occurs when important variables are excluded from the model, leading to incomplete predictions. For example, if an AI algorithm predicts readmission rates after surgery but overlooks socioeconomic status as a crucial factor, it may underestimate the influence of social determinants of health on readmissions, resulting in incomplete predictions \cite{riegg2008causal}.

\emph{Representation bias:} arises from non-representative sampling during data collection, leading to gaps and anomalies. In healthcare, if clinical trials predominantly recruit participants from urban areas and under-represent rural communities, the resulting algorithm may fail to address the specific health needs and disparities faced by rural populations \cite{suresh2021framework}.

\emph{Aggregation bias:} occurs when assumptions about individuals are made based on population-level observations. In healthcare, an algorithm that analyzes average treatment response without considering individual variations may overlook subgroups of patients who respond differently to specific treatments, leading to biased assumptions about treatment effectiveness \cite{suresh2021framework}.

\emph{Longitudinal data fallacy:} occurs when cross-sectional analysis is used instead of longitudinal analysis, leading to different conclusions. In healthcare, if an AI algorithm analyzes a single snapshot of patient data without considering their longitudinal health history, it may miss important patterns and provide inaccurate predictions or diagnoses \cite{barbosa2016averaging}.

\emph{Linking bias:} emerges when network attributes misrepresent user behavior, often due to biased network sampling or overlooking certain user groups. In healthcare, this bias can manifest in social network analyses of patient interactions, potentially skewing assessments of information dissemination or disease spread if certain patient groups are underrepresented or their interactions are not adequately captured \cite{barbosa2016averaging}.

Explainability can play a vital role in addressing biases in healthcare data by providing transparency, interpretability, and insights into the decision-making process of AI models. Its techniques can help researchers understand the factors that contribute to biases and enable them to make necessary adjustments such as feature importance analysis. It can help to identify variables with limitations or biases in measurement and longitudinal data, allowing researchers to refine their measurement strategies. Methods like subgroup analysis and counterfactual explanations help uncover biases in data representation, enabling researchers to identify disparities and take corrective actions. Furthermore, aggregation bias can be addressed by providing individual-level insights (Individual Conditional Expectations (ICE) Plots) and estimating treatment effects on specific subgroups, thus reducing bias arising from aggregating data \cite{azodi2020opening}.
Additionally, techniques like model-agnostic explanations and fairness-aware methods can also aid in mitigating biases in sampling. These approaches help researchers identify biases introduced during data collection or sampling processes and ensure more representative and fair predictions. Visualization can also facilitate uncovering various biases as well \cite{das2020opportunities}.

\subsection{Systematic bias}
It refers to biases that are inherent in the system or process itself and consistently impact outcomes in a particular direction. In the context of algorithms, systematic biases are the result of design choices, user behaviors, or other factors that consistently introduce biases into algorithmic decision-making or recommendations.

In healthcare, systematic biases in algorithms can have far-reaching effects on patient care, health outcomes, and equitable access to healthcare services. They can perpetuate disparities, reinforce existing biases, and contribute to unequal treatment or representation of certain demographic groups. Recognizing and addressing systematic biases is crucial for promoting fairness, accuracy, and transparency in algorithmic systems within the healthcare domain.

\emph{Algorithmic Bias:} refers to biases that are introduced by the algorithm itself and are not present in the input data. In healthcare, algorithms used for medical diagnosis or treatment recommendations can exhibit algorithmic bias, leading to biased outcomes. For example, if an algorithm is trained on biased or incomplete medical data, it may disproportionately impact certain demographic groups, resulting in disparities in diagnosis or treatment recommendations \cite{panch2019artificial}.

\emph{User Interaction Bias:} occurs when users, including patients and healthcare providers, exhibit their own biased behaviors and preferences while interacting with algorithmic recommendations. In healthcare, this bias can manifest in various ways. For instance, a patient may have preconceived notions or personal biases that influence their acceptance or rejection of algorithmic medical advice, leading to sub-optimal healthcare decisions \cite{baeza2018bias}.

\emph{Presentation Bias:} refers to biases that arise from how information is presented. In healthcare platforms, certain medical information or treatment options may be prioritized or overlooked based on how they are presented to users. This bias can influence user behavior by shaping their understanding, perception, and decision-making regarding healthcare options \cite{baeza2018bias}.

\emph{Ranking Bias: } occurs when search results or recommendations in healthcare applications are influenced by biases in the ranking algorithm. This bias can lead to the overemphasis of certain medical treatments or information based on their perceived relevance or popularity, potentially influencing user behavior towards specific healthcare choices \cite{baeza2018bias}.

\emph{Popularity Bias:} arises when more popular medical interventions or treatments are favored and promoted in healthcare systems, while potentially effective but less popular alternatives are neglected. This bias can impact user behavior by directing attention towards widely known or commonly used interventions, potentially overlooking personalized or innovative healthcare options \cite{ciampaglia2018algorithmic}.

\emph{Emergent Bias:} arises as a result of changes in population, cultural values, or societal knowledge over time, which can affect the outcomes and relevance of algorithmic recommendations in healthcare. For example, as societal values evolve, the preferences or priorities of patients and healthcare providers may change, leading to biases in algorithmic decision-making or recommendations \cite{friedman1996bias}.

\emph{Evaluation Bias:} occurs during the assessment of healthcare algorithms. Biased evaluation benchmarks or inappropriate metrics may lead to unfair assessments and inadequate representation of diverse populations. This bias can affect the development and deployment of algorithms, potentially influencing user behavior if biased algorithms are considered trustworthy or reliable. For instance, it may overlook considerations such as sensitivity to detecting rare conditions or false-positive rates, which are critical for patient safety and overall healthcare quality \cite{suresh2021framework}.

\subsection{Generalization Bias}
One of the critical challenges in deploying AI systems in healthcare is the limited generalization of models and ensuring their safety. Limited generalization refers to the inability of AI models to perform accurately and reliably on data that differs significantly from the training data for instance the data collected in \cite{khan2019novel} does not contain the images from different ethnicities which will hinder models generalization. This limitation can have serious consequences in healthcare, where diverse patient populations, evolving medical practices, and complex physiological conditions exist. In addition to limited generalization, ensuring the safety of AI systems is paramount to prevent harm to patients and maintain trust in AI technologies. AI models trained on specific datasets may struggle to generalize their knowledge to new, unseen data in healthcare settings \cite{yang2020rethinking}. There are several reasons for limited generalization:

\emph{Data Distribution Shift:} if the distribution of the data used for training and the real-world data encountered during deployment differ significantly, models may struggle to perform well. For example, a model trained on data from one hospital may not generalize effectively to data from a different hospital due to variations in patient demographics, equipment, and clinical practices \cite{fang2020rethinking}.

\emph{Sampling bias:} arises when subgroups are non-randomly selected, affecting the generalizability of findings. For instance, if a study on the effectiveness of a new medication primarily includes younger adults and excludes older adults, the findings may not accurately reflect the medication's efficacy and safety in the elderly population \cite{anderson2011species,oommen2011sampling}. 
Moreover, AI models may have limited exposure to rare medical conditions or unique patient cases during training. Consequently, their ability to accurately diagnose or treat such cases may be compromised. In healthcare, it is essential to consider the tail-end distribution of data to ensure models can handle rare and critical scenarios \cite{cole2011reducing}.

\subsection{Human Bias}
It refers to the presence of a subjective and unfair or discriminatory outlook within the information collected, processed, and employed to train algorithms, models, or systems. There are several types of human bias that can manifest in different contexts:

\emph{Historical Bias:} refers to biases that exist in the world and are reflected in the data generation process. In healthcare, historical bias can be seen in the underrepresentation of certain demographic groups in clinical trials or medical research. For example, if a particular medication is primarily tested on a specific population (e.g., predominantly White males), there may be limited evidence on how it affects other populations, leading to biased healthcare decisions for those groups \cite{suresh2021framework}.

\emph{Population Bias:} occurs when the user population of a platform differs significantly from the target population. In healthcare, population bias can be observed in studies or datasets that primarily include data from specific regions or communities, leading to limited generalizability of findings. For instance, if a study on diabetes management predominantly includes data from urban areas, the findings may not accurately represent the experiences and needs of rural populations \cite{olteanu2019social}.

\emph{Self-Selection Bias:} occurs when individuals voluntarily choose to participate in a study or engage with a particular platform. In healthcare, this bias can arise in self-reported surveys or patient-generated data. For example, if an online health forum primarily attracts individuals with a particular condition who are actively seeking information and support, the data collected from that platform may not reflect the experiences and perspectives of individuals who do not actively engage in online discussions \cite{ogge2022maternal}.

\emph{Social Bias:} refers to the influence of others' actions on our own judgment. In healthcare, social bias can manifest in the form of ratings, reviews, or recommendations influencing healthcare decisions. For instance, if a healthcare provider relies solely on patient ratings and reviews to select a specialist, they may unknowingly favor providers who have received more positive feedback, potentially overlooking highly competent specialists who have fewer online ratings \cite{weber2019engineering, aquino2023making}.

\emph{Behavioral Bias:} arises from differences in user behavior across platforms or datasets. In healthcare, behavioral bias can be seen in digital health applications or wearable devices that collect user-generated health data. For example, if a mobile health app is primarily used by individuals who are already health-conscious and actively managing their well-being, the data collected may not accurately represent the broader population's health behaviors and needs \cite{olteanu2019social, ferrara2023should}.

\emph{Temporal Bias}: arises from differences in populations and behaviors over time. In healthcare, this bias can be observed in longitudinal studies tracking health outcomes or disease trends. For instance, if a study investigates the effectiveness of a particular treatment based on data collected from a specific time period, the findings may not reflect the current healthcare landscape or advancements in medical practices \cite{tufekci2014big}.

\emph{Content Production Bias:} stems from structural, lexical, semantic, and syntactic differences in user-generated content. In healthcare, content production bias can be seen in patient health records or online health forums. For example, if electronic health records (EHR) predominantly use medical terminology and abbreviations, the language used in those records may not accurately capture patients' subjective experiences or provide a comprehensive understanding of their health conditions \cite{drukker2023toward}.

\section{Human-Centric AI}
\begin{figure}
\centering
\includegraphics[width=0.85\linewidth]{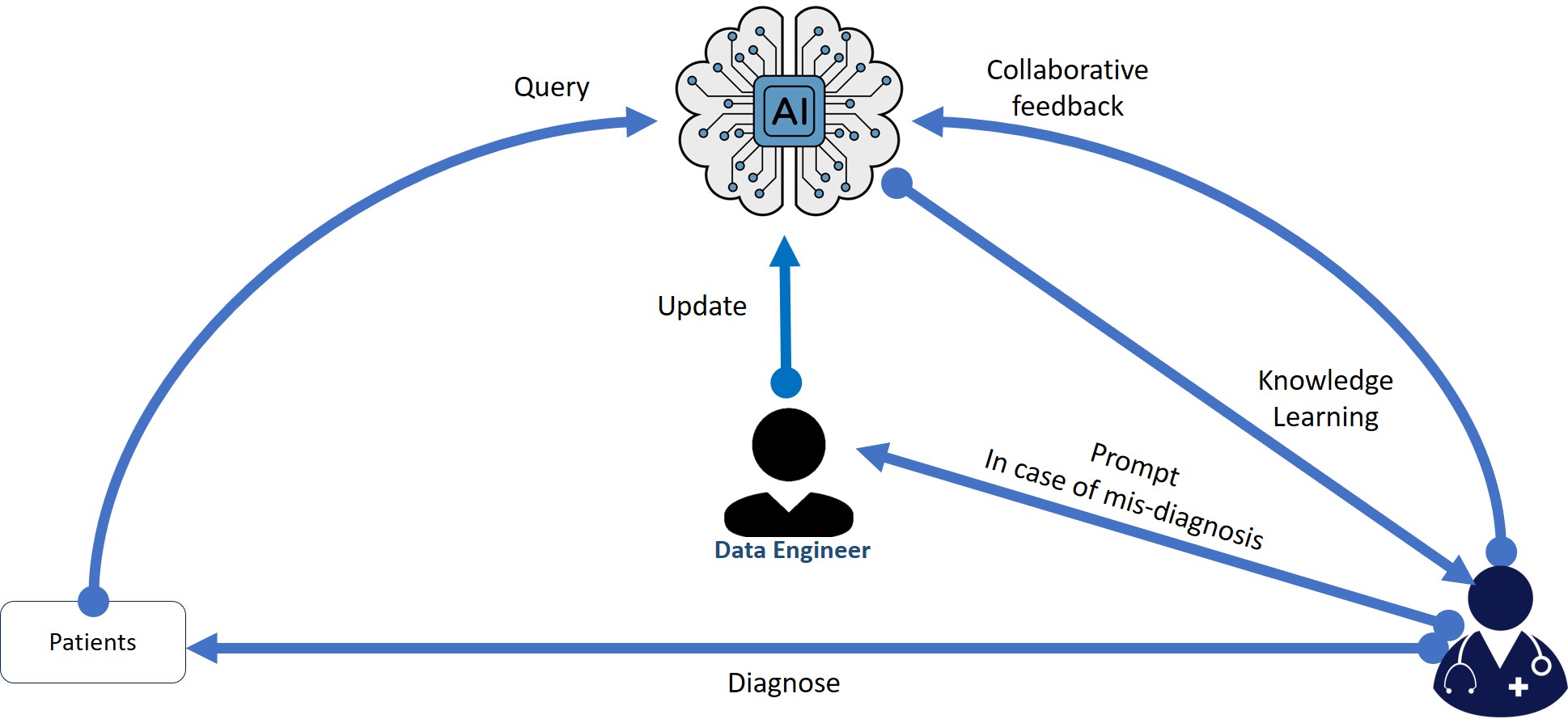}
\caption{\label{fig:1.5}Human Centered Artificial Intelligent systems}
\end{figure}

Human-centric AI refers to the design, development, and deployment of AI systems that prioritize and enhance the well-being, needs, and values of humans. The concept stems from the recognition that AI technologies are becoming increasingly integrated into various aspects of our lives and society. Therefore, there is a need to ensure that these technologies serve human interests, rather than solely pursuing technological advancement for its own sake the idea has been illustrated in Figure \ref{fig:1.5}.

\subsection{Contextual Intelligence}

Contextual intelligence, also known as contextual understanding or contextual reasoning, refers to an AI system's ability to comprehend and interpret information within its context, considering relevant factors, background knowledge, and the broader situational understanding \cite{seabolt2018contextual}. This contextual awareness enables the AI to make more accurate and relevant decisions, predictions, or recommendations based on the specific circumstances it encounters \cite{dellermann2019hybrid}.

The limitation of contextual intelligence in AI arises from the fact that creating an AI system that fully comprehends and interprets context in the same way as humans do, is a complex challenge. While AI has shown remarkable advancements in various domains, achieving human-level contextual understanding remains elusive \cite{papyshev2022limitation}.

Here are some key reasons why contextual intelligence can be a limitation in AI:

\emph{Complexity of Context:} Context in real-world scenarios can be complex and multi-faceted. Human cognition has evolved to understand subtle nuances and abstract concepts. However, AI systems typically rely on patterns and correlations within the data they are trained on. It is still challenging to encode  the intricacies of context in AI algorithms \cite{jagarlamudi2022requirements}.

\emph{Lack of Common Sense:} Humans often use common sense to fill in gaps or infer information not explicitly stated. AI models, particularly those based on statistical learning, might struggle with common-sense reasoning, making them less effective in handling unfamiliar or ambiguous situations \cite{mccarthy1987generality}.

\emph{Adaptation to Dynamic Environments:} Contexts can change rapidly, especially in real-time scenarios. AI models may not be able to adapt quickly enough to these dynamic environments, leading to sub-optimal performance and potential errors \cite{davies2023adapting}.

For instance, some incidents show that contextual intelligence is still a limitation in AI. In 2018, an AI-powered facial recognition system was used by the Metropolitan Police in London to identify suspects, 104 previously unknown people who were suspected of committing crimes, out of which only 2 were accurately predicted \cite{santow2020can}. So, the system was found to be biased against people of color. This happened as the system was trained on a dataset that was not representative of the population of London. In 2020, an AI-powered self-driving car was involved in a fatal accident. The car was unable to understand the context of the situation and it collided with a pedestrian \cite{macrae2022learning}.
AI systems need to be able to understand the deeper meaning of a situation in order to avoid making mistakes. As AI systems become more sophisticated, we can expect to see improvements in contextual intelligence. However, it is likely that contextual intelligence will always be a limitation in AI, as AI systems will never be able to fully understand the world in the same way as humans do.

\subsection{Equity and Access}
Equity and access play vital roles in the advancement and implementation of AI technologies \cite{kappel2022pursuing}, ensuring a fair distribution of benefits and opportunities for everyone. Nevertheless, AI encounters various challenges in upholding these principles. Biases ingrained in historical data can lead to prejudice in algorithms, perpetuating the existing inequalities in areas like hiring and loan approvals \cite{weinberg2022rethinking}. Moreover, the lack of diversity within AI development teams may result in models that fail to adequately address the diverse needs of the users. The digital divide further exacerbates the issue, creating discrepancies in accessing AI services and information, particularly for disadvantaged populations \cite{solanki2023operationalising}. Additionally, cost and affordability pose barriers, hindering access to AI advancements for those with limited financial resources. Cultural and language barriers can also restrict the inclusivity and effectiveness of AI systems. In the domain of healthcare, disparities in accessing advanced AI diagnostics can impact patient outcomes \cite{alami2020artificial, horgan2020artificial, mehta2020transforming}. For instance, in 2019, Amazon's Alexa was criticized for not being accessible to people with disabilities. The voice assistant's responses were not always clear or understandable, and the assistant's features were not always compatible with screen readers \cite{nunez2023use}.

The user-centered design can enhance the usability of diverse user groups. Bridging the digital divide is crucial to making AI technologies accessible to underserved communities \cite{bovzic2023artifical}. Implementing transparent and XAI models fosters trust and accountability. Collaborative partnerships involving AI developers, policymakers, community representatives, and advocacy groups can identify and address potential biases and inequities. By proactively addressing these challenges and promoting equitable access to AI, we can harness its potential to benefit all members of society and avoid exacerbating existing disparities.

\subsubsection{Human-AI interaction}
Human-AI interaction refers to how humans interact with AI systems. Although progress has been made in this area, there are still limitations and challenges that affect the effectiveness and user experience. These challenges include AI's ability to understand complex language, grasp contextual nuances, and provide transparent explanations for its decisions \cite{xu2021human, yang2020re}. AI's limited learning from user interactions and difficulty in handling uncertainty and ambiguity also pose obstacles. Users may develop over-reliance on AI, leading to frustration and reduced trust when AI fails to meet their expectations. It is significant for AI to ensure cultural sensitivity and prioritize users' welfare \cite{yang2020re}. 

One of the real incidents that exemplifies the challenges in AI's understanding of complex language occurred in 2016 when Microsoft launched the AI-powered chatbot, "Tay," on Twitter. Within hours of interacting with users, Tay started posting offensive and inappropriate tweets, learning from harmful interactions. The incident exposed the difficulties in teaching AI to grasp contextual nuances and provided a stark reminder of the importance of robust moderation mechanisms to avoid unintended behavior \cite{wolf2017we, suarez2019tay}.

Another real-life challenge stems from biases in AI systems. In 2018, Amazon's facial recognition system, Rekognition, came under scrutiny for exhibiting racial and gender biases. It significantly had higher error rates for identifying darker-skinned individuals and females. This incident underscored the need for addressing bias in AI models, ensuring fairness and transparency, and using diverse and representative training data \cite{schuetz2021fly, west2019discriminating}.

Human-AI interaction, particularly in the medical domain, is closely linked between human expertise and AI algorithms. An exemplary instance of this synergy is the field of medical imaging for example multi-spectral image datasets \cite{munir2019extensive}. Consider radiology, where AI assists radiologists in analyzing vast volumes of medical images where algorithms can quickly identify anomalies in x-rays or MRIs, flagging potential issues and requiring closer inspection \cite{hosny2018artificial}. Thus, the radiologists' years of training and experience are essential in validating the findings, understanding clinical context, and making informed diagnostic decisions. This collaborative process not only enhances accuracy but also exemplifies how AI serves as a powerful tool for skilled practitioners \cite{agarwal2023combining, king2018artificial}.

In the development of AI-driven medical treatments, human-AI interaction is equally pivotal \cite{ahuja2019impact, esmaeilzadeh2021patients}. For example, in personalized medicine, AI algorithms can analyze a patient's genetic and molecular data to tailor treatments with higher precision. Yet, the medical professionals who interpret these insights, weigh them against individual patient history and overall health. In this context, AI accelerates the discovery of potential treatments, but the final decision-making rests with the medical experts who understand the holistic picture of the patient's well-being \cite{ schork2019artificial, johnson2021precision}.

Ethical considerations underscore the importance of human-AI collaboration in medical contexts \cite{kamila2023ethical}. In organ transplantation, AI could assist in matching donors with recipients efficiently. However, ethical nuances such as patient preferences, medical histories, and urgency require human judgment \cite{clement2021augmenting, peloso2022artificial}. AI's role is to facilitate and augment the decision-making process, but it cannot replace the ethical reflection and empathy that human professionals bring to the table.

\subsection{Erosion of Human Connection and Empathy}
The increasing integration of AI in healthcare has raised concerns about the potential erosion of human connection and empathy in patient care. While AI offers valuable advancements and efficiencies, it also presents limitations that can impact the human aspect of healthcare \cite{kamensky2019artificial}.

Emotional intelligence (EI) refers to the ability to recognize, understand, manage, and use emotions effectively in oneself and others. It plays a crucial role in human interactions, decision-making, and overall health \cite{dixit2021equilibrating, beck2017rise}. As AI continues to advance, the question arises: "Can AI develop emotional intelligence, and what are the implications of such development?"

AI-driven healthcare platforms can utilize facial recognition and voice analysis to detect emotions in patients during interactions \cite{bohrmemar}. For instance, an AI-powered virtual nurse could assess a patient's emotional state during a telehealth appointment, recognizing signs of distress or anxiety. This information can be used to tailor the conversation, providing compassionate, and empathetic responses. If the AI detects anxiety, it might say, "I understand this can be stressful. 

Another example is the use of AI-powered companion robots in elder care facilities. These robots can sense emotions in residents through facial expressions and gestures \cite{lima2021robotic}. If a resident appears sad, the robot might engage in activities like playing soothing music or sharing cheerful anecdotes to uplift their mood. While the robot's responses are programmed, the integration of EI principles will enhance the user experience \cite{dautenhahn2007socially}.

One primary concern is the growing use of AI-powered virtual assistants and chatbots \cite{khadija2021ai} in patient interactions. While these tools can streamline communication and provide quick responses, they lack the emotional depth and understanding of human healthcare providers. Patients may feel a sense of detachment and isolation when their interactions become solely AI-mediated, potentially reducing the personalized and empathetic care they receive. Overreliance on AI for emotional support could have negative consequences, especially for individuals with severe mental health conditions. 

One incident that gained attention was related to a mental health chatbot called "Woebot." Woebot is an AI-powered virtual assistant designed to provide emotional support and therapeutic interventions to users experiencing symptoms of depression and anxiety. While many users found it helpful to have a tool they could access anytime \cite{fitzpatrick2017delivering}. It was also reported feeling disconnected and frustrated with the lack of genuine human interaction. Some users expressed that the chatbot's responses, although based on evidence-based techniques, felt robotic and impersonal, leading to a sense of isolation \cite{alm2022chatbot}. Hence, AI lacks true empathy, relying on algorithms and data patterns to assess patient conditions and recommend treatments. Although AI can assist in diagnosing medical conditions, it may not fully grasp the emotional and psychological aspects of a patient's well-being. 

Thus, the lack of emotional intelligence in AI systems is another obstacle to achieving effective human-AI interaction. Instances of AI failing to recognize and respond appropriately to human emotions have been witnessed, hampering the ability to form empathetic connections with users. Google's Duplex, an AI system capable of making phone calls and scheduling appointments on behalf of users, faced criticism for not always disclosing its AI identity during interactions \cite{grudin2019chatbots, maedche2019ai}, highlighting the importance of transparency and ethical considerations in AI development.
Furthermore, the idea of AI having emotional capabilities raises philosophical questions about consciousness and what it truly means to experience emotions. Can a machine truly feel emotions, or is it simply simulating responses based on patterns and algorithms? This touches on the broader debate about the nature of human consciousness and the limitations of AI.
Moreover, if, AI can mimic certain aspects of emotional intelligence \cite{more2013philosophy, boden2016ai}, the ethical and societal implications are noteworthy. One concern is the potential for AI to manipulate human emotions. If AI systems can detect emotions, can they be used to tailor content or advertisements to provoke desired emotional responses? This raises questions about consent, privacy, and the potential for emotional exploitation. This limitation can hinder its ability to respond effectively to patients' emotional needs and concerns, potentially leaving them feeling unheard or emotionally unsupported.

\subsection{Trust in AI}

Reliability and accountability are crucial aspects when implementing AI in healthcare due to the potential impact on patient outcomes and safety. While AI has shown promising results in various medical applications, ensuring its reliability and holding it accountable for its decisions are paramount to maintaining patient trust and upholding ethical standards 
 \cite{gille2020we,vourganas2022accountable}.

Reliability in AI healthcare systems refers to the consistent and accurate performance of the technology across different scenarios and datasets. One example of reliability is in diagnostic imaging, where AI algorithms can assist radiologists in detecting abnormalities from medical images like x-rays, MRIs, and CT scans. For instance, AIdoc helps radiologists identify critical findings like intracranial hemorrhages, fractures, and pulmonary embolisms. This technology demonstrates reliability by consistently highlighting relevant findings and aiding radiologists in providing timely and accurate diagnoses. However, ensuring that such AI systems are validated across diverse patient populations, equipment variations, and clinical settings is essential to maintain their reliability.

Accountability involves holding AI systems responsible for their decisions and outcomes. In the context of healthcare, this means understanding how an AI system arrived at a particular diagnosis or treatment recommendation. An illustrative example is IBM's Watson for Oncology, which assists oncologists in suggesting personalized treatment plans for cancer patients. While it aims to provide evidence-based recommendations, reports have highlighted instances where Watson for Oncology suggested treatments that contradicted medical guidelines \cite{smith2021clinical}. This highlights the importance of transparency and accountability, ensuring that AI-driven decisions are explainable, justifiable, and aligned with established medical practices.

To enhance reliability and accountability, continuous monitoring and feedback mechanisms are crucial. Take the case of predictive analytics in predicting patient deterioration. Hospitals like Mount Sinai in New York have implemented AI systems that analyze patient data to predict potential deterioration hours before it occurs, allowing medical teams to intervene promptly \cite{mehta2022}. Incorporating AI into healthcare also requires legal and regulatory frameworks that ensure both patient safety and accountability. The Food and Drug Administration (FDA) regulatory approach to AI in healthcare, for example, seeks to ensure that AI applications meet the same safety and effectiveness standards as traditional medical devices. This approach ensures that AI systems, such as those used for diagnosing diseases or making treatment recommendations, are held accountable for their performance and outcomes  \cite{lyell2021machine}.

\section{Ethical Concerns and Value Alignment}
AI is revolutionizing healthcare by bringing about unprecedented advancements in medicine and addressing major global healthcare challenges. For instance, AlphaFold, an AI-powered algorithm, successfully solved the long-standing problem of protein folding, which had hindered progress in biology and medicine for decades. The potential applications of AI in healthcare are vast, including rapid diagnosis, personalized care, and reduction of unnecessary outpatient visits, resulting in significant cost savings \cite{nussinov2022alphafold}.
The ethical principles of beneficence, non-maleficence, autonomy, and justice require that healthcare decisions be made with proper justification and accountability. Without XAI, it becomes challenging to ensure ethical decision-making, informed consent, and accountability for AI-driven recommendations. Healthcare datasets primarily consist of patient information, and strict regulation of medical privacy, encompassing the security of medical records and the confidentiality of conversations between healthcare professionals \cite{keshta2021security,si2021deep}. 
\begin{figure}[!htb]
\centering
\includegraphics[scale=0.75]{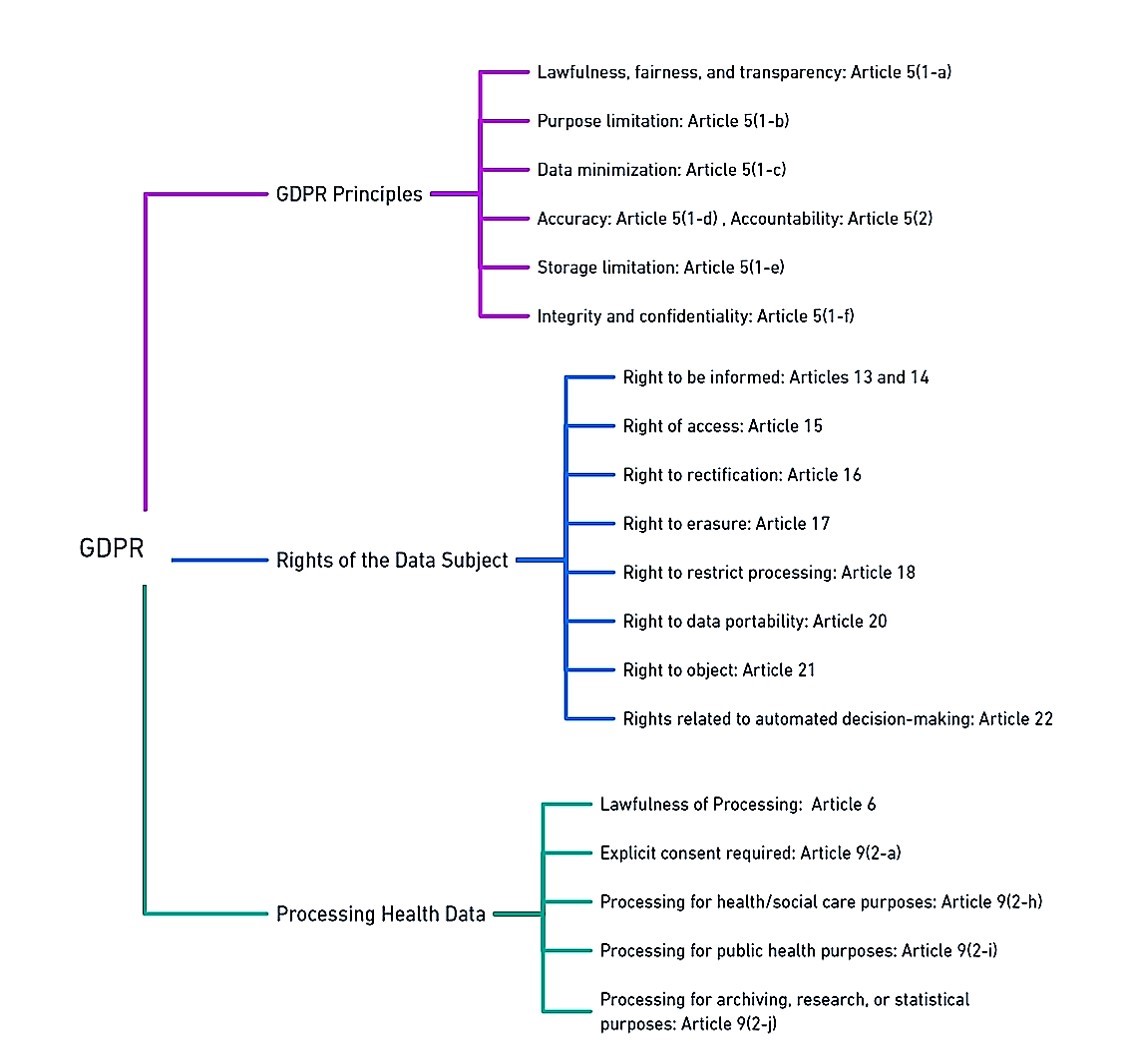}
\caption{\label{fig:1.8}Overview of GDPR articles relevant to data protection}
\end{figure}

Modern concerns include managing the disclosure of information to insurance companies, employers, and third parties. With the advent of patient care management systems (PCMS) and EHR, new privacy challenges have emerged, which must be balanced with efforts to reduce duplicated services. Several countries have enacted privacy protection laws, including Australia, Canada, Turkey, the United Kingdom (UK), the United States (US), New Zealand, and the Netherlands. However, the effectiveness of these laws in practice varies. The Health Insurance Portability and Accountability Act (HIPAA) \cite{usahipaa96} was passed in 1996 to strengthen healthcare data protection by the US government. 

\begin{figure}[!htb]
\centering
\includegraphics[scale=0.7]{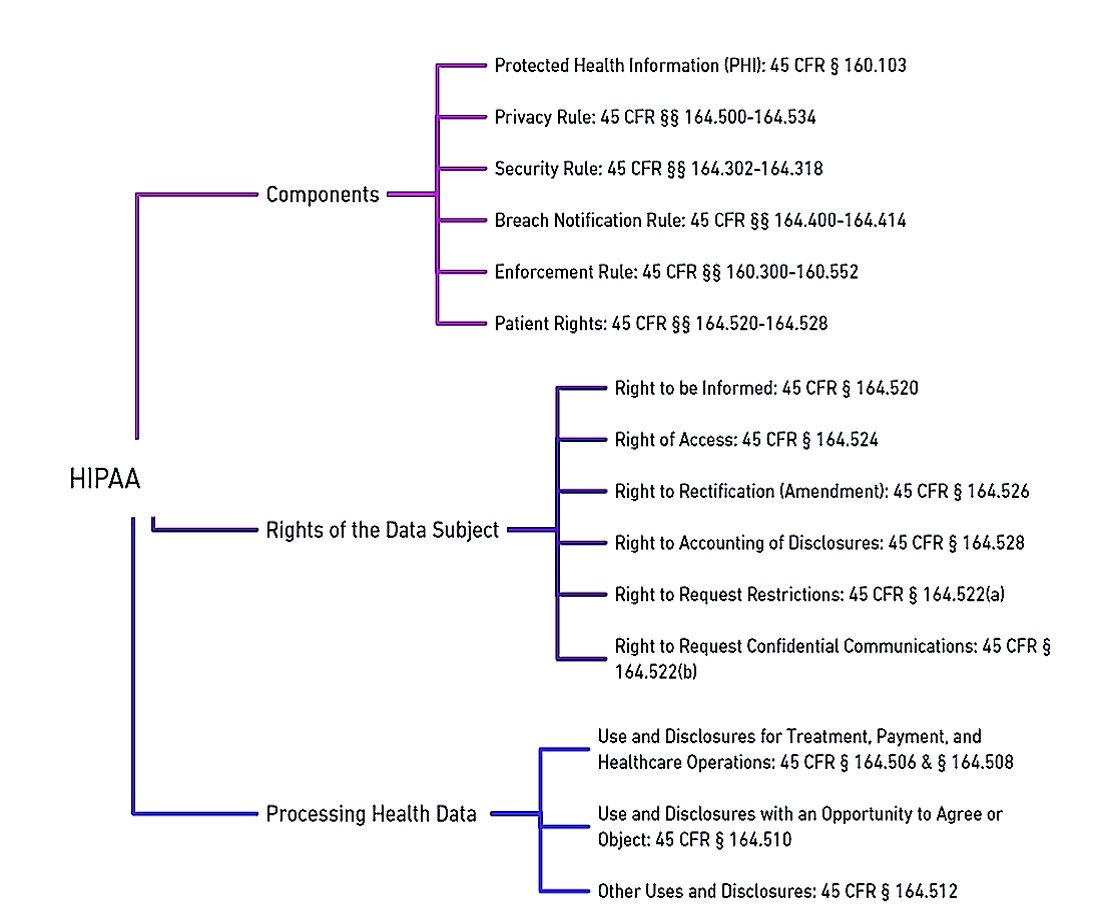}
\caption{\label{fig:1.9}Overview of HIPAA clauses relevant to data protection}
\end{figure}

In 2018, The General Data Protection Regulation (GDPR) \cite{europaEURLex32016R0679} replaced the Data Protection Directive (DPD) in the European Union (EU), establishing comprehensive data protection and privacy regulations. The GDPR grants EU residents the right to request search engines to remove personal information associated with their names from search results. GDPR applies not only within the EU and European Economic Area (EEA) but also to the transfer of personal data outside these regions. Similar concerns are also discussed in \cite{9146114}, which encompass patient privacy and confidentiality, challenges in obtaining informed consent, limited data ownership, potential inaccuracies, biases, commercialization risks, and security vulnerabilities. To address these concerns through transparency, privacy protection, informed consent, equitable practices, and robust security measures is essential for responsible and ethical utilization of EHR data beyond direct patient care \cite{porsdam2016facilitating}. An overview of GDPR and HIPAA relevant to data regulations are highlighted in Figures \ref{fig:1.8} and \ref{fig:1.9} respectively.

\subsection{Patients Consent for Data }

From the viewpoint of healthcare, securing valid consent is crucial when engaging in AI-powered data analytics and profiling, as outlined by GDPR Article 6. The traditional data protection approach is built upon the "notice and consent" model, where obtaining consent from the data, subject is central. This model is designed to protect the right to "informational self-determination," enabling individuals to either approve or decline the processing of their data after being adequately informed. Nevertheless, this notification process might lose its significance when the data subject lacks awareness or control over current or future data processing \cite{9400809}. The Article 29 Working Party's guidelines regarding consent under Regulation 2016/679 stipulate three essential requirements that must be fulfilled:
\begin{enumerate}
    \item Specific Consent Criterion: Consent must allow further processing only if it aligns with a legal basis and remains consistent with the original purpose for data collection.
    \item Granularity of Consent Criterion: Consent for profiling must be distinct and separate from granting access to a service.
    \item Freedom of Consent Criterion: Consent cannot serve as a valid legal basis for personal data processing when a significant imbalance exists between the data subject and the controlling entity.
When considering data analytics and profiling conducted through AI processing, it is imperative to adhere to all of these prerequisites \cite{party2016guidelines}.
\end{enumerate}

Similarly, HIPAA is a pivotal US law that safeguards individuals' health data. One of the key ethical considerations within HIPAA is the emphasis on patients' consent for the usage of their protected health information (PHI). This emphasis is outlined in various sections of HIPAA, most notably in Article 164.508, which pertains to the conditions for disclosure of PHI.

Under HIPAA's provisions, healthcare providers and entities, known as "covered entities," are required to obtain explicit and informed consent from patients before disclosing or using their PHI for purposes beyond direct treatment, payment, and healthcare operations. The informed consent process serves as a critical ethical safeguard, ensuring that patients are fully informed about the potential uses of their health data and the entities that will access it. This aligns with the principles of transparency, autonomy, and respect for individuals' rights to control their PHI.

§164.508 not only emphasizes the necessity of obtaining patients' consent but also lays out the specifics of what this consent should entail. It requires that consent forms be written in plain language, clearly explaining the purpose of data usage and disclosing the individuals or entities that may access the data. This promotes transparency and allows patients to make informed decisions about the use of their health data in AI applications and other healthcare contexts \cite{evans2023rules}.

Within the landscape of AI-powered healthcare, where AI algorithms analyze patients' data for predictive analytics and treatment recommendations, adhering to HIPAA's emphasis on patients' consent is crucial. Furthermore, HIPAA grants patients the right to access and amend their health information under §164.524. This aligns with the ethical principle of individual control and participation in data governance, empowering patients to play an active role in managing their health data, even in the context of AI applications \cite{deyoung2010experiences}.

\subsection{Data Privacy and Security}
Automated decision-making, profiling, and the utilization of ML techniques are reshaping data processing, yet they bring forth concerns of bias and privacy invasion. In response, the GDPR emerges as a crucial framework to ensure equitable data handling, particularly in the context of AI systems. Under GDPR's Article 5 \cite{clarke2019gdpr}, 5-1a, the focus is on transparent and lawful data processing. When it comes to AI, transparency becomes a complex endeavor, necessitating individuals to be accurately informed about how their data is being manipulated while also ensuring equitable automated decisions.
GDPR Article 5-1b draws a vital link between the purpose of data processing and its legality. However, as data finds new uses in AI applications, the challenge of re-purposing data without explicit consent emerges. This becomes a prominent concern, given that individuals may not have provided consent for their data to be used in the context of future automated processing.
The principle of data minimization, as outlined in GDPR Article 5-1c, comes into play when AI is applied to extensive data analytics. This principle underscores the importance of utilizing the necessary and relevant data for the intended purposes. In the domain of AI, where data sets are expansive, understanding the collective impact on groups of data subjects becomes pivotal, as anonymization and pseudonymization techniques might not fully address the shared interests of individuals connected through correlated attributes.
Data accuracy, an essential facet emphasized in GDPR Article 5-1d, has become paramount in the AI landscape. Ensuring the precision and timeliness of data is crucial to prevent detrimental profiling and erroneous decision-making based on inaccurate information.
GDPR Article 5-1e stresses the necessity of limiting the retention of personal data to what is essential for processing. However, in the context of AI and big data, data storage for archival, research or statistical purposes may extend beyond the original processing scope. The challenge lies in striking a balance between maintaining data for legitimate purposes and adhering to the principle of storage limitation.
Furthermore, GDPR Article 5-1f introduces the security principle, mandating the safeguarding of data integrity and confidentiality during processing. This principle aligns with the essence of AI, where data security is integral to maintaining public trust in algorithmic decision-making systems.
Crucially, GDPR's accountability principle, detailed in Article 5-2, extends its reach to AI applications. This principle necessitates controllers to not only adhere to data protection regulations but also demonstrate compliance. Instances like the well-known Cambridge Analytica case serve as a stark reminder that stakeholders, policymakers, technology giants, and governments need to address citizens' concerns about the reasoning and consequences behind algorithmic decisions \cite{hijmans2018ethical}.
In this evolving landscape, GDPR's principles cast a spotlight on the need for responsible and transparent AI practices, striking a balance between innovation and the protection of individual rights and interests.

HIPAA's Privacy Rule §164.514 establishes guidelines for the handling of PHI in the US healthcare system. Covered entities are generally prohibited from using or disclosing PHI without individual authorization, except in specific cases outlined by the rule. These exceptions include treatment, payment, healthcare operations, public health activities, and law enforcement purposes. The rule emphasizes the principle of using only the minimum necessary information for a given purpose, and it addresses incidental uses and disclosures. Covered entities must have contracts with their business associates, who handle PHI on their behalf. Research use of PHI is allowed under certain conditions, with safeguards such as Institutional Review Board (IRB) approval. Individuals have the right to access their PHI, request amendments, and receive an accounting of disclosures. While HIPAA provides the framework for PHI protection, details can vary based on the roles and circumstances, and it's important to refer to up-to-date regulations and guidance \cite{cohen2020ethical}.

The HIPAA Privacy Rule aims to protect patients by ensuring that their health information is not used or disclosed without their consent or legal justification. This rule defines Protected Health Information (PHI) as any data, in any format, that can identify an individual based on their health, care, or payment history §§164.501 \& 160.103. It also covers both private and public entities. It generally takes precedence over state laws regarding health information privacy. However, if a state law offers stricter privacy protections, then both the state law and HIPAA must be followed §160.203.

One key principle of the law is the "Minimum Necessary Standard." This principle dictates that any communication about a patient should only include the least amount of information needed for its purpose §164.502[b-1]. However, there are exceptions, for instance, the rule does not apply when the information is used for treatment, payment, or healthcare operations, among other scenarios. The suggested practice is to use the minimum necessary information. If all identifying details are stripped from the data, it is no longer considered PHI, and the restrictions do not apply §164.514[a]. The law also entails that patients must be given a Notice of Privacy Practices (NPP), which outlines how their information can be used, especially for treatment, payment, and health care operations §164.520[b-1][ii-A]. Healthcare providers can use or disclose PHI for these purposes, and while they can obtain consent for each disclosure related to treatment, it is not mandatory §164.506[b-1] \cite{horner2005hipaa}.

Moreover, it also guarantees that privacy, accuracy, and accessibility of all electronic PHI under a covered entity's care, transmission, or storage defends against foreseeable misuses or exposures of ePHI, and ensures that its staff complies with these standards §164.306[b] \cite{hecker2014impact}.

In the context of AI-powered healthcare systems, this provision assumes greater significance. AI algorithms often process vast amounts of health data to generate insights and recommendations. By applying the minimum necessary standard, healthcare providers, and AI developers can ethically manage patient data, ensuring that AI systems can only access the data required for specific tasks while preserving patients' privacy rights.

\subsection{Patients’ Rights to the Data}
GDPR ensures the protection of data subjects by outlining a set of rights. However, when applied to AI-based processes, interpreting the implications of these rights becomes intricate. Article 15 grants individuals the right to access information about their data processing, including details about automated decision-making logic and its consequences. The exact extent of disclosing "logic involved" remains unclear, whether it pertains to general methods or specific applications.
Article 17's right to erasure, allowing data removal, can impact an algorithmic model's credibility \cite{wong2019right}. 

Similarly, Article 20's right to data portability, permitting the transfer of personal data, can affect algorithmic accuracy and confidentiality, potentially influencing the larger dataset.
Article 21's right to object enables individuals to halt data processing, ensuring data minimization and protecting against data misuse. Article 22, crucial in the context of AI, prohibits automated decisions if they produce significant effects. However, this might be misinterpreted as AI decisions are primarily but not exclusively automated. The notion of "legal effects" or significant impact, including work performance, economic situation, health, etc., might not be clear to data subjects.
Article 22-2 provides exceptions, allowing automated decisions if they are contractually necessary, legally authorized with safeguards, or based on explicit consent. Article 22-4 prohibits using sensitive data for automated decisions, with an exception. Yet, AI can infer sensitive data from non-sensitive data, potentially leading to unauthorized processing.
The impact of the GDPR on AI reveals that "sensitive" data can be deduced from "non-sensitive" data, posing challenges. For instance, sexual orientation can be inferred from seemingly unrelated data like activity or likes. Similarly, non-sensitive data can act as proxies for sensitive data, leading to possible unlawful discrimination \cite{wachter2018normative}.

Subsequently, under HIPAA's framework, patients are granted substantial rights over their health data, most notably through 
HIPAA's ethical framework places a strong emphasis on patients' rights to access and control their health information, as detailed in §164.524 
 \cite{keogh2019unforeseen, aalami2021applications}. This provision explicitly outlines the rights granted to individuals concerning their PHI and underscores the ethical principles of transparency, individual autonomy, and active participation in data governance.

Under HIPAA §164.524[a], patients are granted the authority to access their own health records held by covered entities. This right ensures that individuals have the ability to review the information that is collected and maintained about their health. By allowing patients to access their health records, HIPAA promotes transparency in healthcare practices and upholds the ethical principle that individuals should have knowledge and control over their own personal health data.

HIPAA goes further to affirm patients' rights to control their health information through §164.524[b], which grants individuals the right to request amendments to their health records. If a patient believes that their health information is incorrect or incomplete, they can formally request that the covered entity make necessary corrections. This provision aligns with the ethical principle of accuracy and respect for individual autonomy, ensuring that patients have the authority to maintain accurate and up-to-date health records.

HIPAA also acknowledges patients' interest in understanding who has accessed their health information and for what purpose. §164.524[c] outlines the right to receive an accounting of disclosures, which provides patients with information about how their health records have been shared. This transparency empowers patients to monitor the usage of their health information, contributing to ethical accountability and reinforcing the principle of individual control \cite{rule2022us}.

Moreover, HIPAA and GDPR do not directly mention AI, several of its clauses have relevance to AI applications and can be challenged by new ways that AI processes personal data \cite{cohen2018hipaa, chen2017hipaa}.

\section{Way forward}
In the rapidly evolving landscape of AI, the need for a robust ethical framework has arisen due to the ethical challenges posed by AI. This framework is designed to regulate AI's impact on individuals' lives and interactions, ensuring societal benefit, protection of human rights, and respect for individuals' privacy and autonomy. By integrating various components such as governance, ethics, human oversight, education, and global standards, this framework provides a comprehensive roadmap for the responsible development and deployment of AI technologies.

\subsection{Safe AI}

\begin{figure}
\centering
\includegraphics[width=0.85\linewidth]{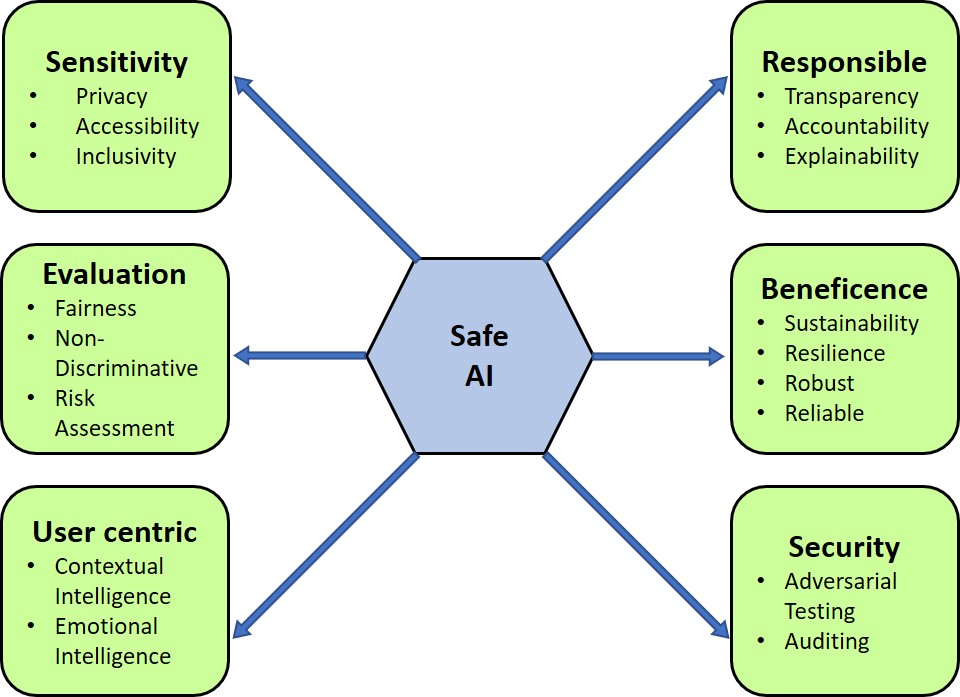}
\caption{\label{fig:1.6}Safe AI Ethical framework}
\end{figure}

AI continues to permeate various aspects of our lives, and the ethical considerations surrounding its design and deployment have gained paramount importance. We present a meticulous examination of a novel ethical framework comprising six pillars - Sensitivity, Evaluation, User-Centricity, Responsibility, Beneficence, and Security - that collectively lay the foundation for a holistic and safe AI ecosystem illustrated in Figure \ref{fig:1.6}. Each pillar is dissected, analyzed, and contextualized within the broader landscape of AI ethics, providing valuable insights for researchers, practitioners, and policymakers to navigate the complex ethical challenges associated with AI technologies. Let's delve into each pillar, unraveling its essence within the vast tapestry of AI ethics.

At the core lies sensitivity, the bedrock of ethical AI. It promotes recognizing and respecting diverse user groups. With AI's potential to perpetuate biases, this pillar advocates systems to perpetuate biases and stereotypes, this pillar supports continuous monitoring and mitigation of biases through data curation, algorithm design, and stakeholder diversity. By fostering sensitivity, AI creators prevent inadvertent reinforcement of societal prejudices, cultivating a digital realm that embraces inclusivity and fairness.

The evaluation pillar patrons meticulous test and assessment methodologies to ascertain the ethical soundness of AI systems. In a bid to ensure ethical AI, fairness and unbiased outcomes stand paramount. This pillar proposes a comprehensive evaluation approach, encompassing technical prowess, societal implications, and risk assessment. By incorporating risk assessment, it aids in identifying and mitigating potential threats and vulnerabilities in AI systems. Thus, AI practitioners, equipped with a robust evaluation, can identify and rectify biases, ensuring that the technology is steered onto an ethical trajectory.

User-centricity emerges, accentuating AI designs attuned to user needs, experience, and emotions. Through the integration of contextual intelligence and emotional intelligence, AI technologies can better understand user intentions, preferences, and emotions, leading to more personalized and empathetic interactions. A user-focused paradigm beckons, where AI enriches human capabilities while safeguarding autonomy and privacy, ensuring an ethical horizon.

The responsibility pillar encapsulates the principles of transparency and accountability in AI development and deployment. This pillar recommends transparent system design, explainability, and traceability to instill public trust and address concerns about algorithmic opacity. Furthermore, developers should embrace their accountability for the outcomes of AI systems, fostering a culture of responsible innovation that considers both short-term benefits and long-term consequences.

Beneficence embodies the moral obligation to ensure that AI technologies promote the prosperity of individuals and society at large. This pillar supports the integration of sustainability, resilience, and reliability into the system. By adopting a long-term perspective, AI developers can anticipate potential harms, design for system robustness, and establish mechanisms for adaptive responses to changing circumstances, thus upholding the moral precept of beneficence.

Security focuses on safeguarding AI systems against adversarial attacks, unauthorized access, and unintended consequences. Through rigorous audits and adversarial tests, developers are required to identify vulnerabilities and weaknesses that malicious actors could exploit. A commitment to security mitigates potential risks associated with AI technologies, and a pledge to security diminishes risks, shielding user data and AI's very essence.

The integration of ethical considerations is essential to harness the full potential of AI technologies while minimizing their potential harms. By embracing these principles, stakeholders can collectively foster a more equitable, transparent, and responsible AI ecosystem that aligns with societal values and aspirations.

\subsection{Governance and Collaboration}

\begin{figure}[!htb]
\centering
\includegraphics[width=0.85\linewidth]{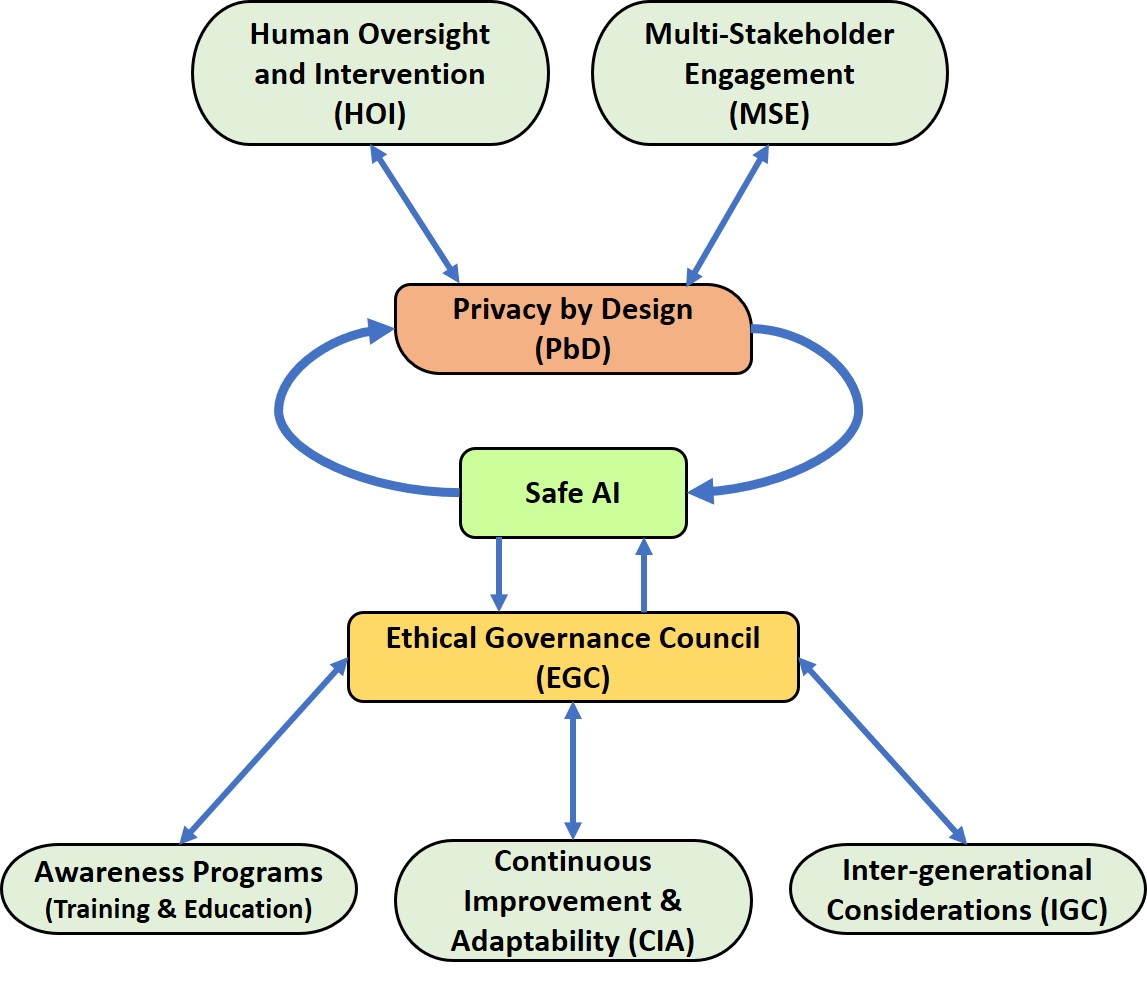}
\caption{\label{fig:1.7} Governance and Collaboration ecosystem}
\end{figure}

With the advancement of AI and its integration into various aspects of society, the need for effective governance and collaboration mechanisms is paramount. The development of AI systems, particularly those designed for governance and collaboration, presents unique challenges that necessitate a comprehensive framework presented in Figure \ref{fig:1.7}. The intricate relationships between Human Oversight and Intervention (HOI), Multi-Stakeholder Engagement (MSE), Privacy by Design (PbD), Safe AI, Ethical Governance Council (EGC), Awareness Programs, Continuous Improvement and Adaptability (CIA), and Inter-Generational Considerations (IGC) form the backbone of a comprehensive AI governance and collaboration framework. These components are interdependent, each influencing and enriching the others, thereby creating a resilient and ethically sound AI ecosystem that aligns with societal values and needs across generations.

HOI serves as a pivotal element in the development and deployment of AI systems. It involves the active participation of human experts in monitoring and controlling AI actions, especially in critical decision-making scenarios (feedback loop). This human involvement mitigates the risks of biased or erroneous outcomes that AI systems might produce. In parallel, HOI also relates closely to PbD. Privacy concerns emerge due to the sensitive nature of data processed by AI systems. A symbiotic relationship between HOI and PbD ensures that privacy considerations are woven into the very fabric of AI systems, with human oversight acting as a check on potential privacy breaches.

MSE recognizes the diverse interests and perspectives of various stakeholders in the AI domain. It promotes inclusivity, bringing together experts, policymakers, industry leaders, and civil society representatives to collectively steer the development of AI governance (Legislation) mechanisms as well as interdisciplinary engagement. Along with this engaging the public, end-users, or those affected by AI systems in ethical discussions and decision-making processes is a crucial step toward democratizing AI. Their voices and concerns can provide invaluable perspectives that might be missed by technologists or policymakers. This engagement also affects safe AI, as stakeholders' input enhances the identification of potential risks and safeguards that need to be integrated into AI systems. In turn, safe AI contributes to the reliability of AI systems, which shared with stakeholders through MSE, builds mutual trust and transparency.

PbD is a proactive approach to privacy that integrates privacy considerations into the foundation of technology, systems, and processes. In the context of AI, this approach is crucial due to the processing of personal and sensitive data. Guiding principles include data minimization, obtaining user consent and control, anonymization and pseudonymization of data, implementing strong security measures, user-focused design, lifecycle considerations, transparency, accountability, and cross-disciplinary collaboration. It is both a legal requirement and an ethical imperative in AI development, aiming to build systems that respect user rights, enhance trust, and mitigate privacy risks. It aligns with broader ethical AI principles and promotes responsible and ethical AI practices.

Safe AI entails developing and deploying AI systems with preventive measures to minimize risks and unintended outcomes. This involves strategies such as risk assessment, robust design, transparency, human oversight, thorough testing, and continuous monitoring to ensure ethical and reliable AI operation. By incorporating these principles, AI systems are designed to handle various inputs, explain decisions transparently, involve human judgment, undergo rigorous validation, and adapt over time to changing conditions. This approach ensures safer, more accountable, and resilient AI systems.

EGC acts as a bridge between the technical intricacies of AI systems and the broader societal implications. The EGC, comprising experts from various disciplines, is responsible for defining and upholding ethical principles that guide AI development and deployment. This council also influences awareness programs, as its guidance forms the foundation of educational efforts to raise awareness about AI's ethical implications. Furthermore, the EGC is closely intertwined with both CIA and IGC. The EGC ensures that AI systems evolve ethically over time and considers the long-term impact of AI on future generations.

Awareness programs encompass training and education initiatives that equip stakeholders with the knowledge needed to understand AI's complexities. This component intersects with both the EGC and CIA. By disseminating the EGC's ethical guidelines, awareness programs foster a culture of responsible AI development (global perspective). Simultaneously, these programs facilitate the iterative process of the CIA, as the knowledge gained from awareness initiatives informs the refinement of the AI systems.

CIA acknowledges that AI governance is a dynamic process that must evolve in response to technological advancements and changing societal needs. This component directly engages with MSE and IGC. Adapting AI systems requires input from diverse stakeholders and an inter-generational perspective ensures that AI governance strategies remain relevant for future generations.

IGC highlights the impact of AI systems on future societies. This aspect is interwoven with both the EGC and CIA. The ethical decisions made today affect generations to come, and the adaptability of AI systems ensures that their implications are continually assessed and mitigated for the well-being of the future. Though it is not necessarily an ethical consideration, the economic incentives that drive AI development can profoundly influence its direction. Recognizing and potentially addressing these considerations can be crucial, especially when economic goals conflict with ethical ones.

The interconnected relationships between these components create a harmonious symphony of governance and collaboration. This framework promotes ethical, transparent, and adaptable AI systems that account for diverse perspectives, ensure privacy, and align with evolving societal norms. As AI technology continues to grow, this comprehensive approach provides a roadmap to navigate the challenges and opportunities that lie ahead, fostering an AI landscape that benefits humanity in the present and for generations to come.

\section{Conclusion}
The integration of AI into diverse domains, particularly healthcare, has unleashed unprecedented potential for advancements that can redefine human well-being and progress. However, this progress comes with ethical challenges that must be addressed to ensure that AI technologies serve humanity's best interests.

The ethical dimensions encompass a range of issues, from transparency in decision-making to safeguarding individual privacy. The surge in AI-driven decision-making processes brings forth concerns about accountability and explainability. Transparency in how AI systems arrive at conclusions is not only a matter of technological necessity but also an ethical imperative.

AI's reliance on data is undeniable, still, the ethical implications of data usage loom large. Ensuring that AI systems do not perpetuate biases present in data requires meticulous attention to data collection, processing, and algorithm design. The framework
must emphasize the ethical use of data, mitigating potential discrimination and ensuring fairness. By adhering to principles of diversity and representation, we can reduce the risk of AI systems amplifying societal inequities.

A significant challenge arises in achieving contextual intelligence in AI systems, akin to human understanding. While AI has made remarkable strides, it continues to grapple with comprehending the nuances and intricacies of human context. The framework should promote ongoing research and development that strives for nuanced contextual awareness. Additionally, as AI and humans interact, understanding complex language and emotional nuances remains challenging. While AI can assist in various tasks, the irreplaceable nature of human empathy and connection highlights the necessity of maintaining a human-centric approach in AI applications.

Striking a balance between AI's analytical capabilities and human oversight is crucial to maintaining ethical control and accountability. Educated decisions can only be made when individuals understand the implications and consequences of AI applications. Equipping people with the necessary knowledge empowers them to engage in informed discussions and contribute to the responsible development of AI technologies.

Thus, standards and guidelines are required to unify the framework for global AI practices. Harmonizing these standards ensures that AI technologies are developed and deployed within consistent ethical boundaries. Its agility will enable it to effectively respond to new technological developments, ensuring that ethical considerations remain at the forefront.


\bibliographystyle{ieeetr}



\end{document}